\documentclass[prl,aps,superscriptaddress,amsmath,amssymb,floatfix,epsf,showpacs,twocolumn]{revtex4-1}
\usepackage{times}
\usepackage{graphicx}
\usepackage{float}
\usepackage{latexsym,amsmath,amssymb,bm,euscript}
\usepackage{color}
\usepackage{subfigure}
\usepackage{epstopdf}
\usepackage[colorlinks=true,linkcolor=blue,citecolor=blue,urlcolor=blue]{hyperref}
\usepackage{hyperref}
\usepackage{ulem}
\usepackage[dvipsnames]{xcolor}
\usepackage{dcolumn}     
\usepackage{slashbox}     
\usepackage{bbm}

\newcommand{\be}[0]{\begin{equation}}
\newcommand{\ee}[0]{\end{equation}}
\newcommand{\ba}[0]{\begin{eqnarray}}
\newcommand{\ea}[0]{\end{eqnarray}}
\newcommand{\mx}[0]{\begin{pmatrix}}
\newcommand{\ex}[0]{\end{pmatrix}}

\newcommand{\kbf}{\ensuremath{\bm k}}
\newcommand{\qbf}{\ensuremath{\bm q}}

\newcommand{\up}[0]{\uparrow}
\newcommand{\dn}[0]{\downarrow}

\begin{document}

\title{Multiorbital spin-triplet pairing and spin resonance in the heavy-fermion superconductor $\mathrm{UTe_2}$
}

\author{Lei Chen$^{\oplus}$}
\affiliation{Department of Physics and Astronomy, Rice Center for Quantum Materials,  Rice University,
Houston, Texas, 77005, USA}

\author{Haoyu Hu$^{\oplus}$}
\affiliation{Department of Physics and Astronomy, Rice Center for Quantum Materials,  Rice University,
Houston, Texas, 77005, USA}

\author{Christopher Lane}
\affiliation{Theoretical Division and Center for Integrated Nanotechnologies, Los Alamos National Laboratory, Los Alamos, New Mexico 87545, USA}

\author{Emilian M.\ Nica}
\affiliation{Department of Physics, Arizona State University, Box 871504, Tempe 85287-1504, AZ, USA}

\author{Jian-Xin Zhu}
\affiliation{Theoretical Division and Center for Integrated Nanotechnologies, Los Alamos National Laboratory, Los Alamos, New Mexico 87545, USA}

\author{Qimiao Si}
\affiliation{Department of Physics and Astronomy, Rice Center for Quantum Materials,  Rice University,
Houston, Texas, 77005, USA}

\begin{abstract}
The heavy-fermion system $\mathrm{UTe_2}$ is a candidate for spin-triplet superconductivity, 
which is of considerable interest to quantum engineering. Among the outstanding issues is the nature of the pairing state.
A recent surprising discovery is the observation of a resonance in the spin excitation spectrum 
at an antiferromagnetic wavevector [C. Duan {\it et al.}, Nature \textbf{600}, 636 (2021)],
which stands in apparent contrast to the ferromagnetic nature of the interactions expected in this system.
We show how the puzzle can be resolved by a multiorbital spin-triplet pairing
constructed from local degrees of freedom. 
Because it does not commute with the kinetic part of the Hamiltonian, the pairing
contains both intra- and inter- band terms in the band basis. We demonstrate that
the intraband pairing component naturally yields a spin resonance at the antiferromagnetic wavevector. 
Our work illustrates how orbital degrees of freedom can enrich
the nature and properties of spin-triplet superconductivity of  strongly-correlated quantum materials.
\end{abstract}


\maketitle

{\it Introduction.~~}
The uranium-based dichalcogenide $\mathrm{UTe_2}$ is a prime candidate for spin-triplet superconductors~\cite{Ran2019_1,Aoki2019,Ran2019_2,Tokunaga2019,Knebel2019,Metz2019,Sundar2019,Jiao2020,Hayes2021,Nakamine2021}.
Because spin-triplet
pairing~\cite{Rice_1995,Kallin_2009,Ramires_ruthenate2019,Huang2019} can lead to
 topological states that host Majorana fermions, which represent a promising
route towards
fault tolerant quantum computing~\cite{Nayak2008}, this system is of extensive current interest.
$\mathrm{UTe_2}$ transitions to a superconducting state below $T_c=1.6$\,K.
It has a strongly-correlated paramagnetic normal state,
with a Kondo temperature of about $70$ K marking the onset of 
heavy-fermion behavior such as 
a highly enhanced effective mass~\cite{Kirchner2020}.
Evidence for the spin-triplet superconductivity includes
the nearly constant Knight shift as observed in 
nuclear magnetic resonance (NMR) measurements~\cite{Ran2019_1,Tokunaga2019,Nakamine2021} across $T_c$,
large upper critical fields exceeding
 the Pauli limit~\cite{Ran2019_1},
and indications of chiral edge states from scanning tunneling microscopy (STM) experiments~\cite{Jiao2020}.
There is also evidence of time-reversal symmetry breaking 
from a non-zero polar Kerr effect~\cite{Hayes2021}.
Analogy with the uranium-based heavy-fermion superconductors $\mathrm{UGe_2}$, $\mathrm{URhGe}$ and $\mathrm{UCoGe}$~\cite{Saxena2000,Aoki2001,Huy07.1}
suggests the natural possibility that ferromagnetic (FM) spin correlations 
are important for superconductivity in  $\mathrm{UTe_2}$.
Early evidence for the role of FM correlations in the normal state comes from the enhanced 
bulk magnetic susceptibility \cite{Ran2019_1} 
as well as by $\mu$SR  measurements~\cite{Sundar2019}. 
Still, the nature of the pairing state
remains enigmatic~\cite{Ishizuka2019,Xu2019,Nevidomskyy20.x,Hu2020,Shishidou2021,Kreisel2021}.

Recently, several experiments~\cite{Thomas2021,Duan2020,Knafo2021,Butch2021,Duan2021,Raymond2021} have suggested
the presence of antiferromagnetic (AF) spin correlations in $\mathrm{UTe_2}$.
 The inelastic neutron scattering (INS) experiments 
reveal spin fluctuations at an AF wavevector, 
 a feature that has been attributed to the Ruderman-Kittel-Kasuya-Yosida (RKKY) interactions between the 
 $\mathrm{U}$ 5$f$ electrons~\cite{Duan2020}. 
Strikingly, measurements below $T_c$ have uncovered a  spin resonance 
near the AF wavevector ~\cite{Duan2021,Raymond2021}.
The observation of an AF spin resonance came as a surprise,
since it has typically been associated with spin-singlet pairing~\cite{Rossat1991,Mook1993,Fong1995,Dai2015}.
In the spin-singlet pairing case, 
the spin resonance has, quite extensively,
 been attributed to collective
excitations of a particle-hole spin composite, at a wave vector that connects 
parts of the Fermi surfaces with a sign change in the superconducting order parameter
 ($\Delta_{k}=-\Delta_{k+q}$) ~\cite{Fong1995,Liu1995.1,Zha1993,Eschrig2006}.
In the case of $\mathrm{UTe_2}$, the central new questions are two-fold: how does spin-triplet pairing 
develop when the 
overall magnetic fluctuations are AF, and how does the spin-triplet pairing in turn
lead to an AF spin resonance?
 
In this work, we advance a resolution to these seemingly contradictory issues.
The key idea, alluded to by some of us in Ref.~\onlinecite{Duan2021}
 and developed here, is that pairing in the strongly-correlated (heavy-fermion) system
$\mathrm{UTe_2}$ 
is to be constructed in terms of its multiple local degrees of freedom that we refer to as orbitals;
these include the sublattice degrees of freedom associated with different
 U-sites in a unit cell.
As a result, the spin-triplet pairing is a matrix not only
in spin space but also in orbital space. 
Reminiscent of a nontrivial class of multiorbital spin-singlet pairing  
states \cite{Nica2017,Nica2021,Pang2018,Smidman2018},
the pairing matrix does not commute with
the kinetic part of the Hamiltonian.
This in turn implies that, in the band basis, the pairing matrix contains both
intra- and inter-band pairing components. We show that, while the overall pairing matrix has the spin-triplet odd-parity 
form,
the intraband pairing component gives rise to a spin resonance.
We achieve an understanding of the spin dynamics 
at the proof-of-principle level, leaving more quantitative 
analyses for 
future studies.
 We note in passing that the interplay between AF correlations and  
 spin-triplet pairing is potentially relevant to other heavy-fermion superconductors,
 including the venerable case of UPt$_3$ \cite{Aeppli1988,Sauls1994,Joynt2002}
  and the more recently-discussed quantum critical systems
 \cite{Nguyen2021,Hu2021.2,Pixley_2015}.

{\it Competing magnetic interactions and microscopic Hamiltonian.~~}
$\mathrm{UTe_2}$ crystallizes in an orthorhombic, centrosymmetric structure, with space group 
$71$ ($Immm$) and two $U$ atoms per unit cell. 
Electronic-structure calculations~\cite{Shick_Fujimori_Pickett_PRB_2021, Schick_Pickett_PRB_2019,Duan2020},
core-level photoelectron spectroscopy~\cite{Fujimori_JPSJ_2021,Fujimori_JPSJ_2019}
and the spin size extracted by inelastic neutron scattering \cite{Duan2021} are consistent with 
the 
U $5f$ electrons 
in a doublet of predominantly $j=5/2$, with $m_{j}= \pm 1/2$.

\begin{figure}[t]
\centering
\includegraphics[width=1\columnwidth]{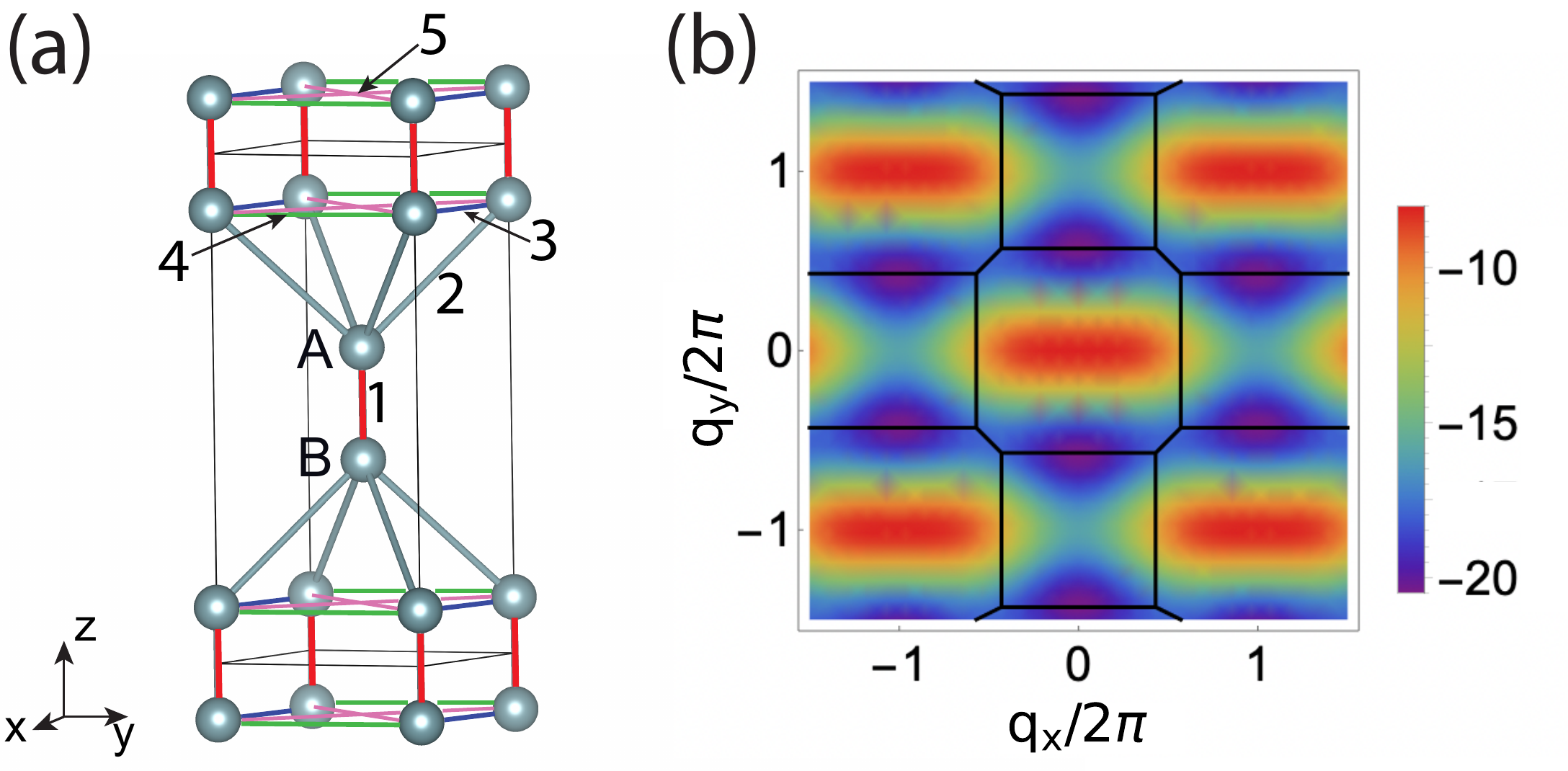}
\caption{ 
(a) The structure of $\mathrm{UTe_2}$. The numbers label the bonds for 
the exchange interactions and kinetic parameters.
The black thin lines mark
 the conventional unit cell, which contains two primitive cells.
(b) 
The exchange interaction eigenvalue, $J_{-}({\qbf})$, plotted within the $(q_x,q_y)$ plane for 
$q_z=0$. It is the most negative 
around the AF wavevector, ${\qbf} \sim \mathrm{\bf Q}$. The color bar marks the values 
in meV.
}
\label{fig:latt}
\end{figure}

We thus consider two Kramers doublets per unit cell.
The effective interactions are labeled by $J_i$'s ($i=1-4$) in Fig.\,\ref{fig:latt}(a). 
We take the exchange interactions to be
\be
\begin{aligned}
H_J
= -\sum_{i} I^{\mu} S^{\mu}_{i,A} S^{\mu}_{i,B} + \sum_{ir,mn}  J_{r}\bm{S}_{i,m} \cdot \bm{S}_{i+r,n}. 
\end{aligned}
\ee
Here, $A/B$ labels the two sublattices [Fig.\,\ref{fig:latt}(a)], 
$i$ marks a unit cell and $S^{\mu}$ is a component of the local spin
for $\mu=x,y,z$. Due to the special role that the intra-``dimer" exchange interaction plays, we refer to
$J_1^{\mu}$ as 
$-I^{\mu}$. In addition, $J_r$ ($r\geq 2$) labels the $r$th-nearest neighbor exchange interaction
that connects $m$ and $n$ sublattices [Fig.~\ref{fig:latt}(a)]. 
We expect to have
significant
$\mu$-dependence 
due to the large spin-orbit coupling (SOC)
that is inherent to the $5f$-electrons of U-ions.
The effect of this spin anisotropy
is evidenced by magnetic susceptibility measurements~\cite{Ran2019_1,Li2021.1}.
In our analysis, only the spin anisotropy of $J_1=-I$ will be important and 
 kept track of. 
In practice, we consider $r$ up to $4$ and have
\begin{eqnarray}
 H_{J}&=& \sum_{\qbf} 
  \begin{pmatrix}
 \bm{S}_{\qbf,A} &  \bm{S}_{\qbf,B}
 \end{pmatrix}
{ J}
(\qbf)
  \begin{pmatrix}
 \bm{S}_{-\qbf,A} \\
 \bm{S}_{-\qbf,B}
 \end{pmatrix} 
\end{eqnarray}
with
\be
\begin{aligned}
{J}(\qbf) = &[2\,J_3 \,T_3 (\qbf) 
+2\,J_4\,
T_4(\qbf)]\,  \tau_0  \\
& + [-I+4\,J_2\,
T_{2,1}(\qbf)] \, \tau_x
+4\,J_2\,
T_{2,2}(\qbf) \, \tau_y,
\end{aligned}
\ee
where the Pauli matrices $\tau_i$ are defined in the orbital space,
while
$T_{2,1}  (\qbf) = \cos(q_x/2) \cos(q_y/2)\cos(q_z/2)$, 
$T_{2,2}  (\qbf) = \cos(q_x/2) \cos(q_y/2)\sin(q_z/2) $, 
$T_3 (\qbf)  = \cos(q_x)$, and 
$T_4 (\qbf)  = \cos(q_y)$.
The INS experiments~\cite{Duan2021,Knafo2021} 
allow for the inference of 
FM
$I$ and $J_3$ and 
AF $J_2$ and $J_4$ 
 (see Supplementary Materials).

The U-U distance 
within a dimer is smaller than all the other U-U distances, and we expect 
$I$ to be the strongest.
We have performed first-principles calculations within density functional theory~\cite{Lane2021} (see also
Ref.~\onlinecite{Xu2019}).
Our calculations give average values of 
$-J_1 \equiv I
\lesssim 
20$ meV and $J_2 \sim 0.75$ meV.
Estimates of $J_{3,4}$ are more demanding as they require a larger unit cell with five magnetic configurations 
considered on an equal footing.
Based on these results and the Kondo temperature scale,
we set 
$I^x=15$ $\mathrm{meV}$,
$J_2=1$ $\mathrm{meV}$, $J_3=-0.5$ $\mathrm{meV}$, and
$J_4=2$ $\mathrm{meV}$. 
The largest negative eigenvalue
of the exchange interaction matrix is:
\begin{equation}
    \begin{aligned}
    J_{-}(\qbf)
    =& -I + 4J_2 \cos \frac{q_x}{2}\cos \frac{q_y}{2}
    +2J_3 \cos q_x + 2J_4 \cos q_y \, ,
    \end{aligned}
\end{equation}
for $q_z=0$.
Importantly, the ferromagnetic intra-dimer interaction $J_1 \equiv -I$ is 
featureless in wavevector space.
Instead, the inter-unit-cell AF
interactions are responsible for the minimum of 
 $   J_{-}(\qbf)$ near
$\qbf
\sim \mathrm{\bf Q}= (0,\pi)$,
as  shown in 
Fig.\,\ref{fig:latt}(b).

The Hamiltonian is that of a Kondo lattice, with the $5f$-electrons occupying 
 $\Gamma_5$ Kramers doublets on the U sites that are 
Kondo-coupled to the $spd$-electrons and, in addition, interact with each other 
through $H_J$. 
For our proof-of-principle purpose,
we adopt an effective description 
that is based 
exclusively 
on effective $f$-electrons and incorporate $H_J$ and a kinetic part $H_{kin}$ (see below).

{\it Spin-triplet pairing -- orbital basis.~~}
The strongly-correlated nature of the $5f$ electrons suggests the construction of pairing based 
on the local degrees of freedom~\cite{Duan2021}.
Given that the FM intra-dimer interaction $I$ is the strongest,
we consider the dominant pairing to be between the $5f$ electrons on a dimer, 
as introduced in Ref.\,\onlinecite{Hu2020} and
subsequently in Ref.\,\onlinecite{Shishidou2021}. 
An odd-parity pairing can be
constructed in terms of 
 two fermions with opposite parity 
[the even and odd parity combinations of the fermions
at the $A$ and $B$ sites, $c_{e/o} =
(c_A \pm c_B)/\sqrt{2}$, where $c_{e}\,(c_o)$ is even\,(odd)]
and  are, equivalently, written as follows:
\begin{eqnarray}
\Delta_{x} &=& \delta_x   \tau_y \otimes \sigma_x \otimes \gamma_1 \nonumber\\
\Delta_{y} &=& \delta_y   \tau_y \otimes \sigma_y \otimes \gamma_1 \nonumber\\
\Delta_{z} &=& \delta_z   \tau_y \otimes \sigma_z \otimes \gamma_1 \, .
\end{eqnarray}
Here, $\delta_x,\delta_y,\delta_z$ denote the pairing amplitudes
whereas $\sigma_i$ and $\gamma_i$ are respectively Pauli matrices in spin and Nambu
spaces.
A generic 
spin triplet 
pairing can be characterized by the vector $\vec{d}=(\delta_x, \delta_y,\delta_z)$.

In $\mathrm{UTe_2}$, the magnetic susceptibility shows extreme anisotropy~\cite{Ran2019_1}.
In the normal state, the
magnetic susceptibilities satisfy $\chi_x\gg \chi_z>\chi_y$, which is consistent with
 the hierarchy of exchange interaction, 
$I^x\gg I^z>I^y$.
In a triplet superconductor, the direction of the dominant 
exchange interaction promotes superconducting pairing with a perpendicular spin component~\cite{Leggett1975}
(see Supplementary Materials).
Accordingly, the energetically favored pairing component is 
the
$\Delta_y$
channel in the $B_{3u}$ irreducible representation of the $D_{2h}$ point group. 
We denote its overall amplitude to be 
$\delta$, 
and set $\delta_x=\delta_z=0$; extension to including 
these other pairing channels is discussed below.

{\it Spin-triplet pairing -- band basis.~~}
The Bogoliubov de Gennes (BdG) Hamiltonian is now written as 
\begin{equation}
\label{eq:hamorb}
H = \sum_{\kbf} \Psi_{\kbf}^{\dagger}\, [\, H_{kin}(\kbf) + \Delta(\kbf) \, ]\, \Psi_{\kbf} \, ,
\end{equation}
where $\Psi_{\kbf}^{\dagger} = \left ( c^{\dagger}_{\kbf,m,\sigma}, c_{-\kbf,n,\sigma'}(i\sigma_y) \right )$ 
is the Nambu spinor, and
$m,n=A/B$
are the orbital (sublattice) indices.

\begin{figure}[t!]
\centering
\includegraphics[width=1\columnwidth]{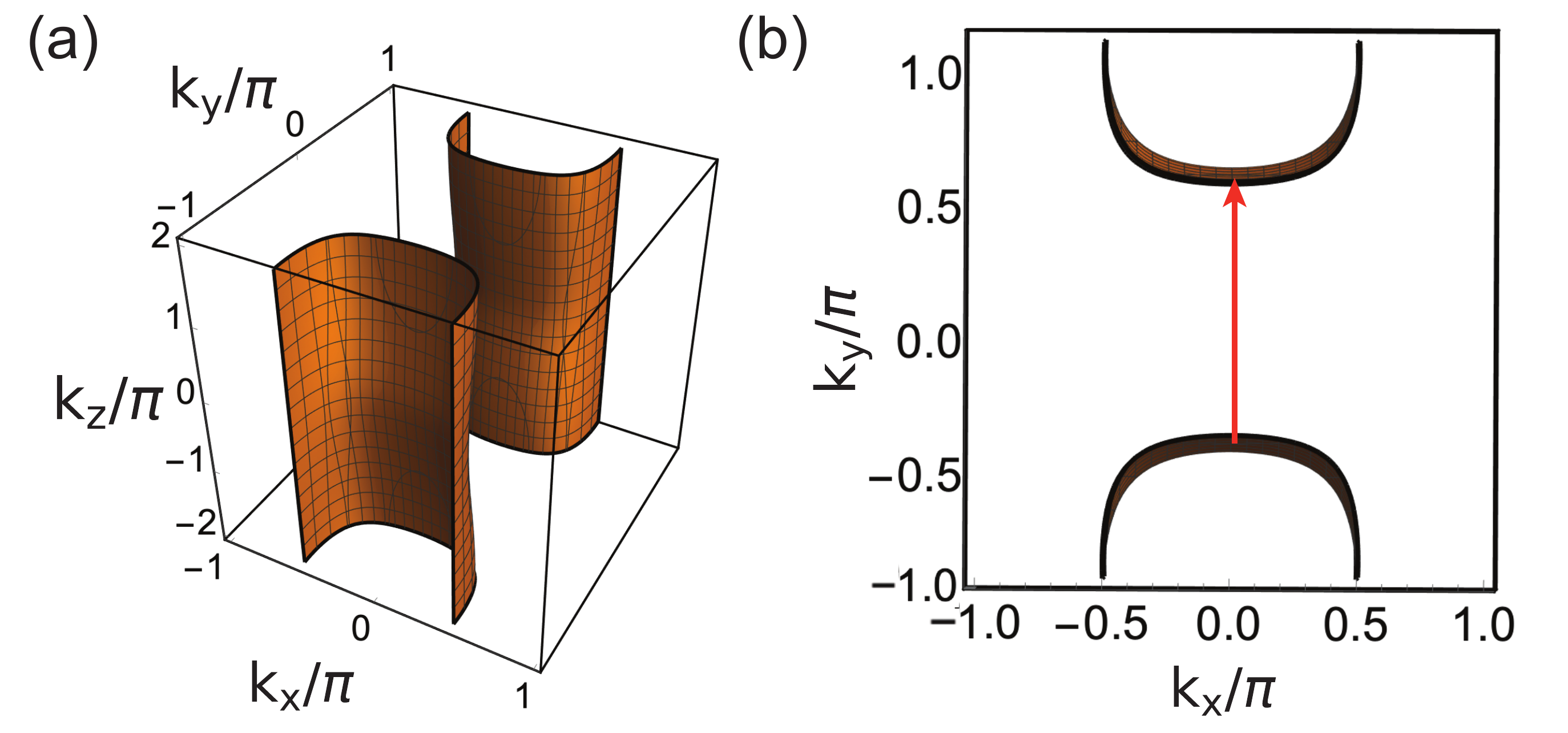}
\caption{ (a) The 3D Fermi surface and (b) the corresponding top view, with the 
spanning wavevector (red arrow) that is close to $\mathrm{\bf Q}$.
}
\label{fig:fs}
\end{figure}

We incorporate the hopping terms $t_r$ 
following the same numbering scheme of the bonds as for the 
exchange interactions, and also allow for $t_5$ defined on the diagonal counterpart of the bonds $3$ and $4$
[c.f., Fig.\,\ref{fig:latt}(a)].  The kinetic part
then has the following form
\be
\begin{aligned}
H_{kin}
= \sum_{i} t_1 c^{\dagger}_{i,A} c_{i,B} +  \sum_{ir,mn} t_r c^{\dagger}_{i,m} c_{i+r,n}. 
\end{aligned}
\ee
In momentum space (see Supplementary Materials for the definition of Fourier transformation), 
$H_{kin}$
becomes:
\begin{equation}
\label{eq:kin}
\begin{aligned}
    H_{kin}(\kbf) = & \xi_0(\kbf) \, \tau_0 + \xi_x(\kbf) \, \tau_x + \xi_{y}(\kbf) \, \tau_y \\
    = &[2\, t_3 \, T_3  (\kbf) + 2\, t_4 \, T_4 (\kbf) + 4 \, t_5 \, T_5 (\kbf) ] 
    \, \tau_0  \\
    & + [t_1 + 4\, t_2 \,T_{2,1}  (\kbf) ] \, \tau_x
    + 4 \, t_2 \, T_{2,2} (\kbf) \, \tau_y \, , 
 \end{aligned}
\end{equation}
where 
$\sigma_{i}$
are 
the usual Pauli matrices 
 in 
spin space,
and $T_5 (\kbf) =  \cos(k_x)\cos(k_y)$.
We note that $\xi_{0,x}(\kbf) = \xi_{0,x}(-\kbf)$ and $\xi_{y}(\kbf) = -\xi_{y}(-\kbf)$. 
Guided by the DMFT calculations of the electronic structure~\cite{Duan2020,Miao2020} 
and the Kondo temperature,
 we take $t_1=-0.95$ $\mathrm{meV}$, $4t_2=1$ $\mathrm{meV}$,
 $2t_3 =-2$ $\mathrm{meV}$,  $2t_4 = 2$ $\mathrm{meV}$, $4 t_{5}=6$ $\mathrm{meV}$,
 and a chemical potential $\mu=3$ $\mathrm{meV}$. 
 The Fermi surface is shown in Fig.~\ref{fig:fs}(a,b),
 which has a quasi- two-dimensional (2D)
  shape with a Fermi pocket surrounding the $(0,\pi,k_z)$ line.
[A suitable variation in $\mu$ can lead to the coexistence of another Fermi pocket 
 surrounding the $(\pi,0,k_z)$ line.]
For these parameters, the total bandwidth is $W = 19$ $\mathrm{meV}$; 
it is comparable to the strength of the exchange interaction [c.f., Fig.\,\ref{fig:latt}(b)], and 
each is about the Kondo temperature
up to the typical conversion factor from temperature to energy
(roughly $3$).

\begin{figure}[t!]
\centering
\includegraphics[width=1\columnwidth]{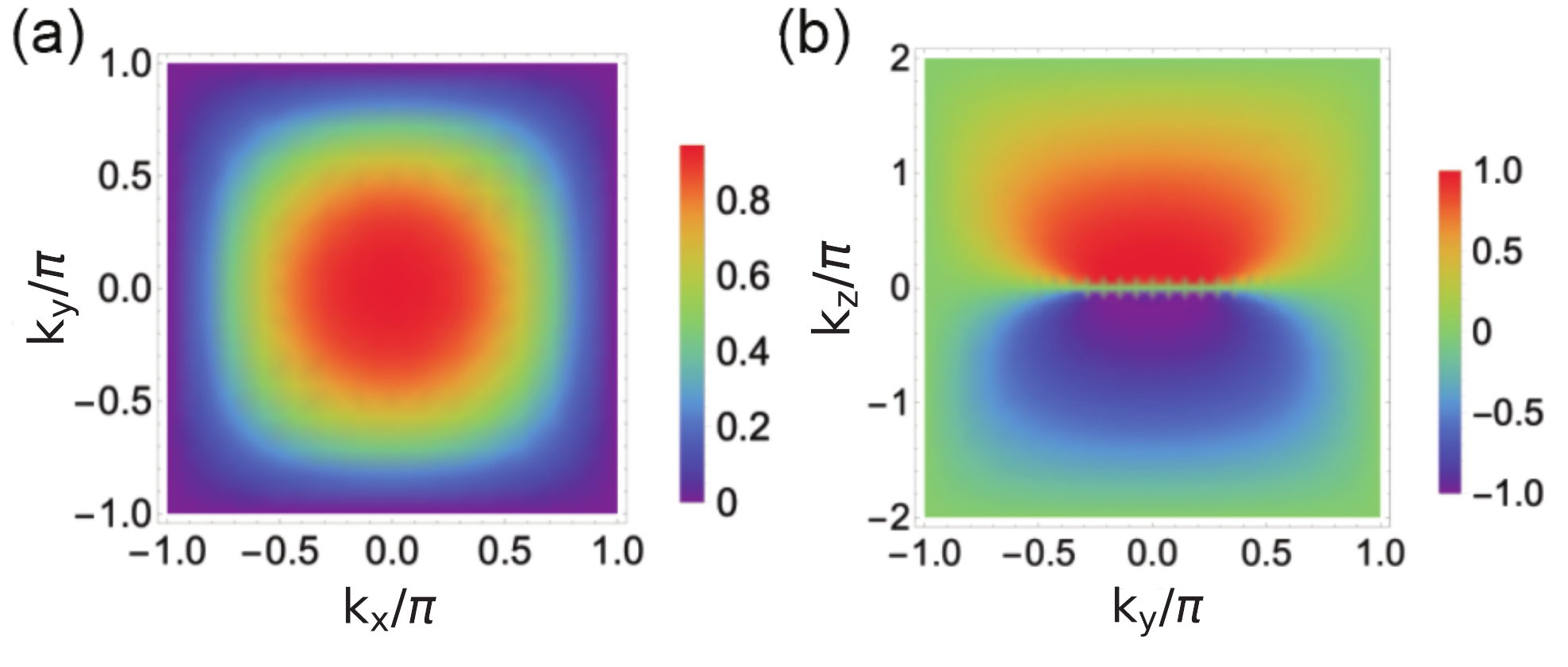}
\caption{Schematic illustration of the
 intraband pairing  ($d_{1}(\kbf)/\delta$)
within (a) the $k_z=\pi$-plane and (b) 
the  $k_x=0$-plane.
}
\label{fig:gap}
\end{figure}

We can now project the pairing onto the band basis.
Given that the kinetic and pairing parts do not commute 
with each other, the two cannot be simultaneously diagonalized. 
The pairing part must contain both intra and interband components. 
To see this, we first introduce a transformation that diagonalizes the kinetic energy:
\begin{equation}
U^{\dagger}_{\kbf}   H_{kin}(\kbf) U_{\kbf}  = \mx
\epsilon_{-}(\kbf) & 0 \\
0 & \epsilon_{+}(\kbf)
\ex \otimes \sigma_0
\end{equation}
where the band energy $\epsilon_{\pm}(\kbf) =\xi_{0} \pm \sqrt{\xi_{x}^2+\xi_{y}^2}$. The explicit form of $U_{\kbf}$ is
given in the Supplemental Materials.
The pairing in the band basis takes the form
\begin{equation}
\Delta(\kbf) = U_{\kbf}^{\dagger} \Delta_y U_{-\kbf} = d_1(\kbf)\alpha_3 \otimes \sigma_2 + d_2(\kbf) \alpha_2 \otimes \sigma_2 \, .
\end{equation}
Here, $\alpha_2$ and $\alpha_3$ are Pauli matrices in the band basis, and the associated 
 inter- and intra- band pairing components are:
\begin{equation}
\begin{aligned}
d_1(\kbf) =& \delta \,
\xi_y(\kbf)/\sqrt{\xi_x^2(\kbf)+\xi_y^2(\kbf)}, \\
d_2(\kbf) =& \delta \, 
\xi_x(\kbf) / \sqrt{\xi_x^2(\kbf)+\xi_y^2(\kbf)}\, . 
\end{aligned}
\end{equation}
The band-diagonal $\alpha_3$ and band-off-diagonal $\alpha_2$ pairing components have $p_z$ and $s$ form factors, respectively. As illustrated in Fig,~\ref{fig:gap}, the intraband gap functions vanish in the $k_z=0,\pm 2\pi$ plane. 

The full BdG spectrum can be obtained by diagonalizing
 Eq.~\ref{eq:hamorb} (see Supplemental Materials), which yields
 \begin{equation}
  \begin{aligned}
  &E_{\pm}(\kbf)   \\
  & = \sqrt{\left(\sqrt{\xi_0^2(\kbf)+\delta^2\sin^2\phi(\kbf)} \pm B(\kbf) \right)^2 + \delta^2(1-\sin^2 \phi(\kbf)) }
  \end{aligned}
  \end{equation} 
 with $B(\kbf) = \sqrt{\xi_x^2(\kbf) + \xi_y^2(\kbf)}$, and  $\sin \phi(\kbf) = \frac{\xi_x(\kbf)}{B(\kbf)}$. 
On the Fermi surface,
$\xi_0^2(\kbf) = \xi_x^2(\kbf) +  \xi_y^2(\kbf)$, and there is a full gap.
 This feature can be modified when the other spin-triplet pairing components are included (see below),
but our conclusion about the AF spin-resonance will be robust.

{\it Spin-triplet pairing and spin resonance.~~}
The orbital-dependent
 spin susceptibility is given by
\be
\chi_{\alpha\gamma}(\qbf, i\omega) = \sum_{\lambda} [\chi^{0}(\qbf,i\omega)]_{\alpha\lambda}[\mathbb{I}+ J(\qbf)[\chi^{0}(\qbf,i\omega)]]_{\lambda \gamma}^{-1} , 
\label{eq:sus}
\ee
where $\alpha$, $\beta$, $\gamma$ and $\lambda$ denote the orbitals.
For $q_z=0$ of our interest, the total spin susceptibility is given by
\ba 
\label{eq:fm_afm_qz=0}
\chi^\mu(\bm{q}) 
=2 [ \chi^\mu_{AA}(\bm{q})
+\chi^{\mu}_{AB}(\bm{q}) ] \, .
\ea
As noted before,
 $\mu=x,y,z$ denotes the spin directions.

Unlike singlet superconducting states,
here 
$\chi^\parallel$ and $\chi^\perp$,
the spin susceptibility parallel and perpendicular to the 
$\vec{d}$ vector respectively, have different superconducting 
coherence factors~\cite{Dirk2001,Maier2008} (see Supplementary Materials). 
Because of the highly anisotropic nature of the exchange interaction $I$, 
the dominating susceptibility is along the $\mu=x$ direction, 
for which we can expect a coherence factor favoring a spin resonance 
to be of the same sign
at $\kbf$ and $\kbf + \mathrm{\bf Q}$.
We will therefore calculate 
$\chi^x$ in the following (unless otherwise specified).

\begin{figure}[t!]
\centering
\includegraphics[width=1\columnwidth]{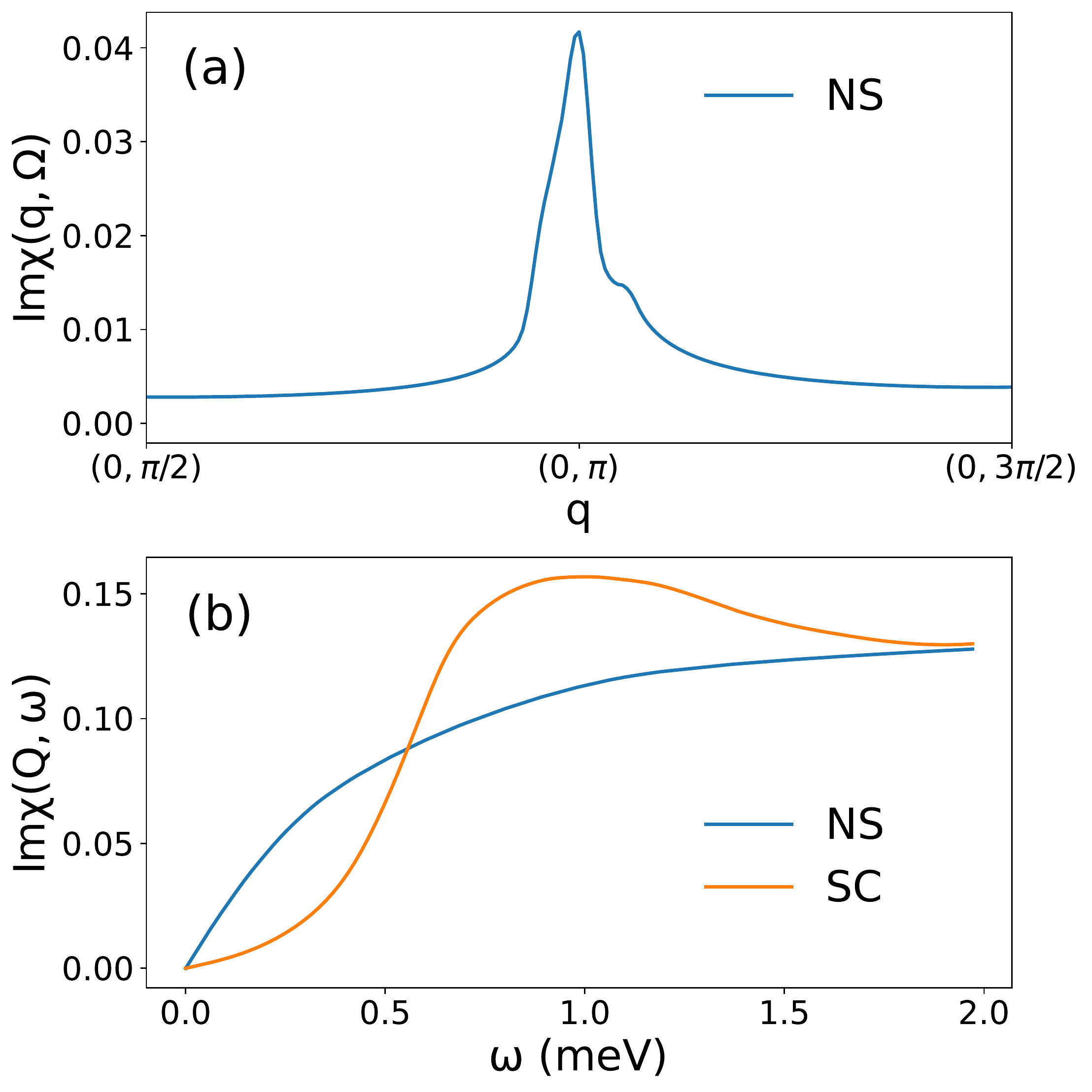}
\caption{ 
(a) The imaginary part of the dynamical spin susceptibility v.s. $q$ in the normal state (NS) with frequency $\omega=0.25\mathrm{meV}$
and (b) of the dynamical spin susceptibility in both the superconducting states (SC) and normal states, at $\mathrm{\bf Q}=(0,\pi)$, 
as a function of frequency ($\omega$, normalized by $\Delta$, the maximal pairing gap on the Fermi surface).
}
\label{fig:res}
\end{figure}

We are now in position to present our results on the spin dynamics.
We first consider the results for the normal state (NS),
obtained from Eq.~\ref{eq:fm_afm_qz=0}
with $\delta=0$.
Fig.~\ref{fig:res}(a) shows the ${\qbf}$-dependence of the imaginary
part of the dynamical spin susceptibility
at a fixed
$\omega= 0.21$ $\mathrm{meV}$, along a
momentum 
path from $(0,\frac{\pi}{2})$ to $(0, \frac{3\pi}{2})$. 
It exhibits an AF peak at wavevectors 
around  $\mathrm{\bf Q} = (0,\pi)$,
which is near where $J_{-}(\qbf)$ is the most negative and
also is close to connecting the Fermi 
sheets.  

In the superconducting phase, we first  solve the pairing 
function self-consistently to 
show the development of spin-triplet pairing.
This is described in the Supplementary Materials, where
we indeed find the $\Delta_y$ pairing. With the model parameters
described earlier, the
 maximal value of the gap close to the Fermi surface is found to be
$\Delta=0.44\mathrm{meV}$.
 
Fig.~\ref{fig:res}(b) shows the energy dependence of $\mathrm{Im} \chi ({\bf Q}, \omega)$,
 the imaginary part of the $J({\qbf})$-enhanced dynamical spin susceptibility at the 
wavevector $\mathrm{\bf Q} = (0,\pi)$,
in the superconducting state. (The bare dynamical spin susceptibility is given in the Supplementary Materials.)
A spin resonance is clearly seen
around $\omega_{res}\approx 1$ $\mathrm{meV}$.
The resonance
 can be regarded as a collective excitation
 from single-particle and single-hole states,
  for all the different $k_z$,
 that are located across the $q_y$ direction.
The $k_z$ variation  of  the intraband pairing component [Fig.~\ref{fig:gap}(b)]
contributes to the broadening of  the resonance.

One of our main conclusions
 is the existence of a spin resonance at the  AF wavevector.
We find the ratio
 $\omega_{res}/\Delta $ to be larger than what is typical (that is significantly less than $2$) 
 for 
the usual 
 quasi-2D (spin-singlet)
superconductors~\cite{Eschrig2006}.
However, the precise value of this ratio
depends on details 
of the interactions and Fermi surface.

Our results naturally provide an overall understanding of the salient experimental-motivated issues
outline earlier. First, the calculated dynamical
spin susceptibility in the normal state [Fig.~\ref{fig:res}(a)] 
realizes the AF fluctuations that have been observed 
by 
INS experiments~\cite{Duan2020,Knafo2021}.
This is further illustrated in Fig.~\ref{fig:Jcmp} (Supplementary Materials),
where the theoretically determined {\qbf}-structure is shown to be comparable with
 the momentum pattern observed 
experimentally. 
Second, by incorporating the underlying
 local degrees 
of freedom, we have shown that such AF correlations promote 
the experimentally-evidenced spin-triplet pairing.
Third, we have demonstrated that the spin-triplet pairing gives rise to an AF spin resonance.
Our results [Fig.~\ref{fig:res}(b)] capture the salient features of the experimental observation,
as illustrated in the Supplementary Materials (Fig.~\ref{fig:res_cmp}).
Finally, the SOC-induced spin-anisotropic interactions built into our model capture the feature 
that the dominant magnetic fluctuations appear in the $x$-direction, as observed in the 
INS measurements~\cite{Duan2021}.

{\it Discussion.~~}
Several remarks are in order. 
First, 
the importance of the local degrees of freedom and the associated multiorbital form of spin-triplet pairing reflect the underlying
strong correlations of the system and are to be contrasted with weak-coupling approaches.

Second,
a main focus of our 
work is a proof-of-principle demonstration that,
when multiorbital pairing is considered,
 the spin-triplet nature of the superconducting state in $\mathrm{UTe_2}$ can naturally be reconciled with
 the AF correlations in the normal state and 
 the emergence of an AF spin resonance 
 in the superconducting state.
This sets the stage for extensions of the model. For example, the pairing function can be generalized to the 
non-unitary case, which can i) break the time-reversal symmetry; and ii) lead to a 
BdG spectrum that not only depends on the usual 
$| \vec{d}|^{2}$ gap term,  but also on an additional splitting proportional to $|i \vec{d}^{*} \times  \vec{d}|$~\cite{Sigrist1991}.
While the nodal structure of the pairing state will be affected, 
 the linkage between the 
spin-triplet pairing and AF spin resonance is expected to be robust. 
In the same vein, our work also sets the stage for addressing the quantitative 
aspects of the spin resonance, such as the resonance energy versus the superconducting pairing gap.
For the latter, we note that the multiplicity of the 5$f$ electrons in $\mathrm{UTe_2}$ may well lead to 
orbital-selective correlations, as has been observed in other U-based heavy-fermion systems~\cite{Chen19.1,Giannakis2019}.
Thus, there 
could very well be multiple pairing gaps that remain to be experimentally probed.
 
Third, the matrix form for the spin-triplet pairing we have considered bears some conceptual 
similarity to its spin-singlet counterpart in the contexts of both the iron-based
superconductors~\cite{Nica2017,Nica2021} and $\mathrm{CeCu_2Si_2}$~\cite{Pang2018,Smidman2018,Amorese20}.
Our work therefore highlights the notion that the emerging conceptual framework for multiorbital superconducting pairing,
which is anchored by the iron-based superconductors~\cite{Yu2014.1,Yin2014.1,Ong2016,Si16.1},
may apply to a variety of correlated systems.

{\it Conclusion.~~}
To summarize, we have highlighted the importance of the matrix structure associated 
with the local degrees of freedom of the heavy-fermion
superconductor $\mathrm{UTe_2}$. This structure allows us to provide a natural understanding of
some outstanding issues
that have been raised for this system. 
The puzzling questions include  how
spin-triplet pairing can develop when the overall magnetic fluctuations are antiferromagnetic,
and how the spin-triplet pairing  can in turn lead to an antiferromagnetic spin resonance.
In our theory, the spin-triplet pairing is enabled by the dominant intra-dimer ferromagnetic
 interactions along with the inter-unit-cell antiferromagnetic interactions. 
 The latter not only are responsible for the observed antiferromagnetic fluctuations but also 
 make them react to the onset of superconductivity.
We expect that the new insight our work reveals, {\it viz.} the local degrees of freedom can 
qualitatively enrich
the nature and properties of spin-triplet superconductivity, will be relevant to a variety of other 
strongly-correlated quantum materials.

{\it Acknowledgement.---} 
We thank Ryan Baumbach, Nick Butch, Sheng Ran, Brian Maple, Chandan Setty, Yi-feng Yang
  and, particularly,
Pengcheng Dai and Chunruo Duan for useful discussions.
Work at Rice was primarily supported
 by the 
U.S. Department of Energy, Office of Science,
Basic Energy Sciences, under Award No.\ DE-SC0018197
and additionally supported by
the Robert A.\ Welch Foundation Grant No.\ C-1411.
The majority of the computational calculations have been performed on the
Shared University Grid at Rice funded by NSF under Grant EIA-0216467, a partnership between
Rice University, Sun Microsystems, and Sigma Solutions, Inc., the Big-Data Private-Cloud 
Research Cyberinfrastructure MRI-award funded by NSF under Grant No. CNS-1338099 and by
Rice University, and the Extreme Science and Engineering Discovery Environment (XSEDE) by NSF
under Grant No. DMR160057. 
Work at Los Alamos was supported by LANL LDRD Program, UC Laboratory Fees Research Program 
(Grant Number: LFR-20-653926), 
and in part by Center for Integrated Nanotechnologies.
One of us
(Q.S.) acknowledges the hospitality of the Aspen Center for
Physics, which is supported by NSF grant No. PHY-1607611.
\\

\noindent
$\oplus$ These authors contributed equally to this work.

\bibliography{ute2}
\newpage

\onecolumngrid 
\section{Supplementary Materials}

\section{Comparing our results with INS experiments}

\begin{figure}[b!]
\centering
\includegraphics[width=0.9\columnwidth]{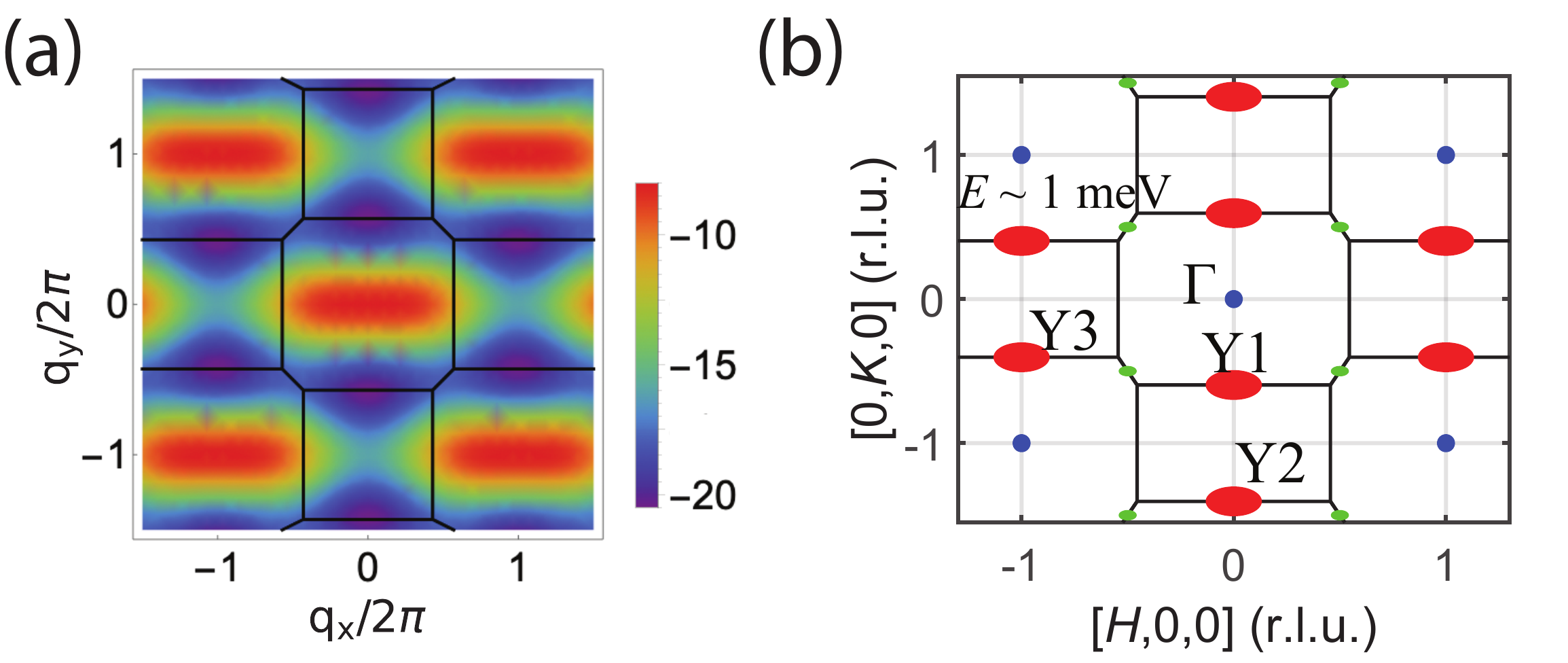}
\caption{ 
(a) The exchange interaction eigenvalue, $J_-(\qbf)$, plotted in the $(q_x,q_y)$ plane for $q_z=0$ 
[Fig.~\ref{fig:latt}(b) of the main text].
(b) The schematic plot of the INS pattern in the $[H, K, 0]$ plane 
that are marked as red ellipses [adapted from Ref.~\onlinecite{Duan2021}].
The minima of  $J_-(\qbf)$ in (a) are consistent with the spin excitation patterns observed in (b).
In both (a) and (b), the Brillouin zone boundaries are marked by the solid black lines.
}
\label{fig:Jcmp}
\end{figure}

In this section, we compare our theoretical results with the INS measurements of Ref.~\cite{Duan2021}.
One comparison is in the momentum domain. As shown in Fig.~\ref{fig:Jcmp}(a) [i.e., Fig.~\ref{fig:latt}(b) of the main text],
 the minima of the exchange interaction are located about 
 the same positions 
 where the AF spin excitations are observed in the INS experiments:
 They appear at BZ boundaries along the $q_y$ direction, as marked 
 by the red ellipses in Fig.~\ref{fig:Jcmp}(b)~\cite{Duan2021}.
 
\begin{figure}[b!]
\centering
\includegraphics[width=0.9\columnwidth]{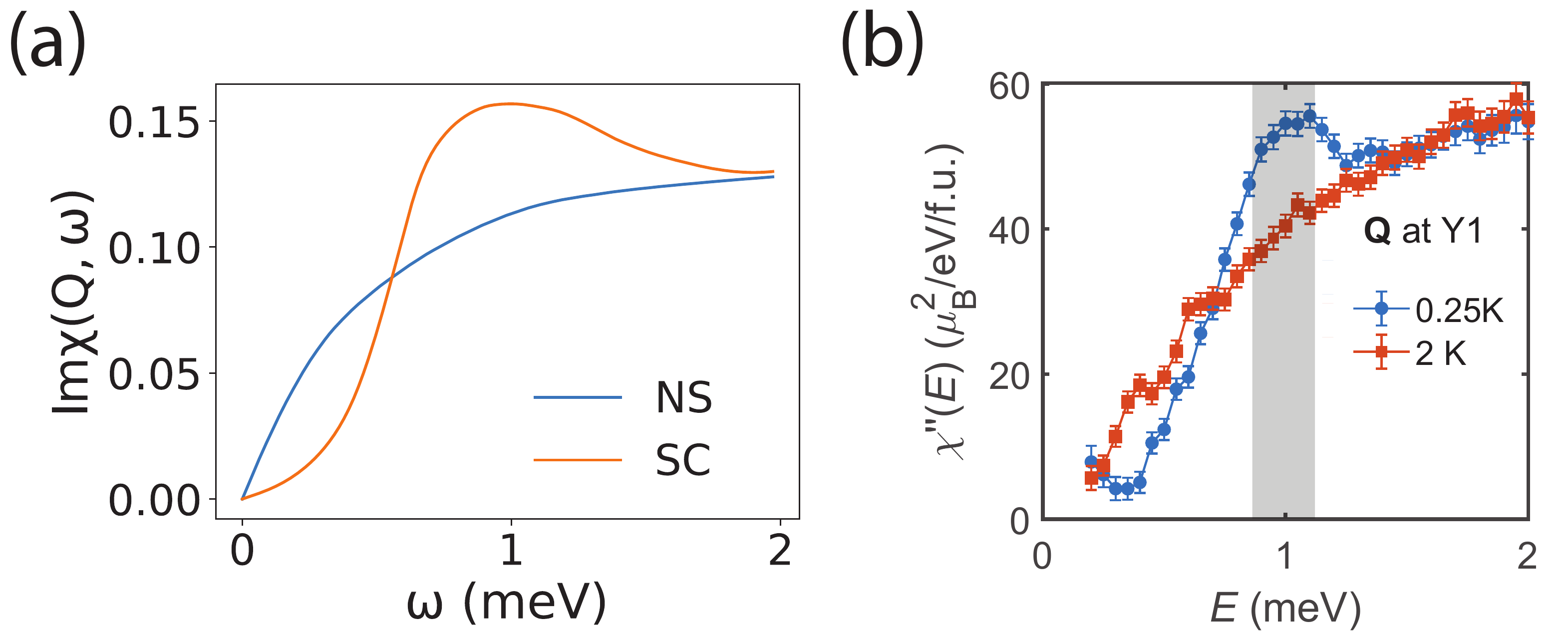}
\caption{ 
(a) The imaginary part of the dynamical spin susceptibility as a function of energy
in both the superconducting (SC) and normal (NS) states [ [Fig.~\ref{fig:res}(b) of the main text]. 
(b) The spin excitations observed at $K=0.59, H=0$ v.s. energy at
$T=0.25$ $\mathrm{K}$ (SC) and $2$ $\mathrm{K}$ (NS) [adapted from Ref.~\onlinecite{Duan2021}]. 
The theoretical results capture the basic features of the 
experimental observations.
}
\label{fig:res_cmp}
\end{figure}

 The other comparison is in the frequency domain. The spin resonance from our calculations in the superconducting state, 
presented in Fig.~\ref{fig:res_cmp}(a) [i.e., Fig.~\ref{fig:res}(b) of the main text], 
captures the basic features of the spin resonance observed in the INS experiments, which is 
displayed in Fig.~\ref{fig:res_cmp}(b)~\cite{Duan2021}.

\section{Pairing in the band basis}

The kinetic term [Eq.~\ref{eq:kin}] in the orbital basis $\Phi_{\kbf} = (A_{\up}(\kbf), A_{\dn}(\kbf), B_{\up}(\kbf), B_{\up}(\kbf))$ [$j_{\sigma}=c_{j,\sigma}$, with $j=A,B$ for short-hand notation] can be diagonalized by a unitary matrix
\begin{eqnarray}
    U_{\kbf} = \frac{1}{\sqrt{2}}\mx
    0 & -u_{\kbf} & 0 & u_{\kbf}\\
    -u_{\kbf} & 0 &u_{\kbf}&0\\
    0 & 1 & 0 &1 \\
    1 & 0 & 1 & 0
    \ex,
\end{eqnarray}
where $u_{\kbf} = \frac{\sqrt{\xi_x^2(\kbf)+\xi_{y}^2(\kbf)}}{\xi_{x}(\kbf)+i\xi_{y}(\kbf)}$. 

The corresponding band dispersions are
\begin{equation}
\epsilon_{\pm}(\kbf) =  \xi_0(\kbf) \pm \sqrt{\xi_{x}^{2}(\kbf) + \xi_{y}^2(\kbf)}.
\end{equation}
Each band is doubly degenerate due to the spin (of the Kramers doublet). We define the annihilation operators of lower and upper bands as $a$ and $b$, respectively,
 and let $\phi(\kbf) = (a_{k,\up}, a_{k,\dn}, b_{k,\up}, b_{k,\dn})^{T}$

The BdG Hamiltonian can be expressed in the Nambu basis $\psi(\kbf) = (a_{k,\sigma}, b_{k,\sigma},  \mathcal{T}a^{\dagger}_{k,\sigma}{\mathcal{T}^{-1}}, \mathcal{T}b^{\dagger}_{k,\sigma}{\mathcal{T}^{-1}})^{T}$, with $\mathcal{T}$ the time-reversal operator, as
\begin{eqnarray}
\label{eq:hamband}
H_{BdG}(\kbf) = \mx
E_{k} & \Delta_k \\
\Delta_k^{*} & -E_{-k} 
\ex,
\end{eqnarray}
where 
\begin{eqnarray}
E_{k} &&= U_{\kbf}^{\dagger} H_{kin} U_{\kbf}=
\mx
\epsilon_-(\kbf) & 0 \\
 0 &   \epsilon_+(\kbf)
 \ex \otimes \sigma_0, \\ \nonumber
&&= 
\mx
 \xi_0(\kbf)- \sqrt{\xi_{x}^{2}(\kbf) + \xi_{y}^2(\kbf)} & 0 \\
 0 &   \xi_0(\kbf)+ \sqrt{\xi_{x}^{2}(\kbf) + \xi_{y}^2(\kbf)}
 \ex \otimes \sigma_0,
\end{eqnarray}
and 
\begin{eqnarray}
    \Delta_k = U^{\dagger}_{\kbf} \Delta_yU_{-\kbf} = \delta \mx
    0 &-id_1 & 0 & d_2\\
    id_1 & 0 & -d_2& 0 \\
    0 & -d_2 & 0 & id_1 \\
    d_2 & 0 & -id_1 & 0
    \ex,
\end{eqnarray}
with $d_1=\frac{\xi_y}{\sqrt{\xi_{x}^2+\xi_{y}^2}}$ and $d_2=\frac{\xi_{x}}{\sqrt{\xi_{x}^2+\xi_{y}^2}}$. The intraband form factor $d_1$ is  $p_{z}$-wave, and the interband form factor $d_2$ is $s$-wave.
  The dispersion of BdG Hamiltonian is
  \begin{equation}
  \begin{aligned}
  E_{\pm}(\kbf) &= \sqrt{\delta^2+\xi_0^2(\kbf)+\xi_x^2(\kbf) + \xi_y^2(\kbf) \pm 2\sqrt{\delta^2\xi_{x}^2(\kbf)+\xi_0^2(\kbf)\xi_x^2(\kbf) +\xi_0^2(\kbf)\xi_y^2(\kbf)}} \\
  \end{aligned}
  \end{equation} 
 where $E_-$ ($E_{+}$) corresponds to the band 
 crossing (above) the Fermi energy. They can be further simplified as
  \begin{equation}
  \begin{aligned}
  E_{\pm}(\kbf)  & = \sqrt{ \delta^2+\xi_0^2(\kbf)+\xi_x^2(\kbf) + \xi_y^2(\kbf) \pm 2\sqrt{\xi_x^2(\kbf)+\xi_y^2(\kbf) } \sqrt{\xi_0^2(\kbf)+\delta^2 \sin^2 \phi(\kbf)}} \\
  & = \sqrt{\left(\sqrt{\xi_0^2(\kbf)+\delta^2\sin^2\phi(\kbf)} \pm \sqrt{\xi_x^2(\kbf) + \xi_y^2(\kbf)}\right)^2 + \delta^2(1-\sin ^2\phi(\kbf)) }
  \end{aligned}
  \end{equation} 
 with $\sin \phi(\kbf) = \frac{\xi_x(\kbf)}{\sqrt{\xi_x^2(\kbf)+\xi_y^2(\kbf)}}$. $E_{-}(\kbf)=0$ requires both squared terms to be 0. The second term equals $0$ in $k_z=0,2\pi$ planes where $\sin^2\phi_k=1$. The first term goes to 0 when
 \begin{equation}
\delta^{2} = \xi_{x}^{2}(\kbf)+ \xi_y^2(\kbf) - \xi_0^2(\kbf).
 \end{equation}
The Fermi surface is specified by 
 $\xi_{0}-\sqrt{\xi_x^2+\xi_y^2}=0$.

Finally, we show the explicit expressions that connect the fermion operators in the orbital and band bases:
\begin{equation}
\begin{aligned}
    A_{\up} &= \frac{1}{\sqrt{2}} \left( -u_k a_{\dn} + u_{k}b_{\dn}\right)
     \\
    A_{\up} &= \frac{1}{\sqrt{2}} \left( -u_k a_{\dn} + u_{k}b_{\dn}\right)
     \\
    B_{\up} &= \frac{1}{\sqrt{2}} \left( a_{\dn} + b_{\dn}\right) 
     \\
    B_{\dn} &= \frac{1}{\sqrt{2}} \left(a_{\dn} + b_{\dn}\right) \, .
\end{aligned}
    \label{eq:project}
\end{equation}

\section{self-consistent determination of the spin-triplet pairing function}
We start from the anisotropic spin-exchange interactions (see main text), 
$I_x \gg I_z >I_y$, and thus consider 
\begin{equation}
\label{eq:mf}
\begin{aligned}
\hat{H}_{ex} = & \sum_i  -I^x S^x_{i,A} S^x_{i,B}  -I^y S^y_{i,A} S^y_{i,B} -I^z S^z_{i,A} S^z_{i,B} \\
= \sum_i& \frac{-I_x+I_y-I_z}{4} \frac{1}{4} \left( \Phi_i^{\dagger} [i\tau_y \sigma_0] \Phi_i^{*} \right)  \left( \Phi_i^{T} [-i\tau_y \sigma_0] \Phi_i \right) \\
& + \frac{I_x-I_y-I_z}{4} \frac{1}{4} \left( \Phi_i^{\dagger} [-\tau_y \sigma_z] \Phi_i^{*} \right)  \left( \Phi_i^{T} [\tau_y \sigma_z]\Phi_i \right) \\
& + \frac{-I_x-I_y+I_z}{4} \frac{1}{4} \left( \Phi_i^{\dagger} [\tau_y \sigma_x] \Phi_i^{*} \right)  \left( \Phi_i^{T} [-\tau_y \sigma_x]\Phi_i \right) \\
& + \frac{I_x+I_y+I_z}{4} \frac{1}{4} \left( \Phi_i^{\dagger} [i\tau_y \sigma_y] \Phi_i^{*} \right)  \left( \Phi_i^{T} [-i\tau_y \sigma_y]\Phi_i \right), 
\end{aligned}
\end{equation}
where $\Phi_{i} = \left( A_{\up}, A_{\dn}, B_{\up}, B_{\dn}\right)_i^{T}$. 
The first three terms correspond to triplet pairing while the last term is the singlet pairing. 
The dominance of $I_x$ (together with $I_z > I_y$)
 implies that we can focus on the first term.  We define
\begin{equation}
\delta_y = \langle B_{\up}A_{\up}+B_{\dn}A_{\dn}-A_{\up}B_{\up}-A_{\dn}B_{\dn} \rangle \, ,
\end{equation}
and perform Hubbard-Stratonovich transformation to Eq.~\ref{eq:mf}. The effective Hamiltonian for our self-consistent analysis becomes
\begin{equation}
H_{eff}
= \sum_{\kbf}\frac{1}{2} \mx 
H_{kin}(\kbf) & -\Delta(\kbf) \\
-\Delta^{\dagger}(\kbf) & -H_{kin}(-\kbf) 
\ex - \frac{I_{eff}}{16} \delta_{y}^*\delta_y,
\end{equation}
and the gap function $\Delta(\kbf)=\frac{I_{eff}\delta_y}{8}i\tau_y\sigma_0$, with $I_{eff} = I_x-I_y+I_z$. 
The pairing amplitude can now be determined self-consistently. We 
find $\delta=0.6\mathrm{meV}$. The maximal value of the corresponding gap close to the Fermi surface is
$\Delta=0.44\mathrm{meV}$.

\section{Spin resonance in a single-band spin-triplet superconductor}
For a single-band spin-triplet $ \Delta_k \sigma^y$ pairing, the bare spin susceptibilities along three directions take the forms
\begin{eqnarray}
\chi^{0,x}(\qbf,i\omega) &=& \frac{1}{4} \sum_k \frac{1}{2}\left(1 - \frac{\epsilon_k\epsilon_{k+q}-\Delta_k\Delta_{k+q}}{E_{k}E_{k+q}} \right)\left( \frac{1-f(E_k)-f(E_{k+q})}{i\omega+E_{k}+E_{k+q}} + \frac{1-f(E_k)-f(E_{k+q})}{-i\omega+E_{k}+E_{k+q}}\right) \\
\chi^{0,y}(\qbf,i\omega) &=& \frac{1}{4} \sum_k \frac{1}{2}\left(1 - \frac{\epsilon_k\epsilon_{k+q}+\Delta_k\Delta_{k+q}}{E_{k}E_{k+q}} \right)\left( \frac{1-f(E_k)-f(E_{k+q})}{i\omega+E_{k}+E_{k+q}} + \frac{1-f(E_k)-f(E_{k+q})}{-i\omega+E_{k}+E_{k+q}}\right) \\
\chi^{0,z}(\qbf,i\omega) &=& \frac{1}{4} \sum_k \frac{1}{2}\left(1 - \frac{\epsilon_k\epsilon_{k+q}-\Delta_k\Delta_{k+q}}{E_{k}E_{k+q}} \right)\left( \frac{1-f(E_k)-f(E_{k+q})}{i\omega+E_{k}+E_{k+q}} + \frac{1-f(E_k)-f(E_{k+q})}{-i\omega+E_{k}+E_{k+q}}\right) 
\end{eqnarray}
where $\epsilon_k$ is the dispersion in the normal state, $\Delta_k$ is the gap function 
and $E_{k} = \sqrt{\epsilon_k^2+\Delta_k^2}$ is the BdG spectrum.
We notice that, the resonance depends on both the direction of the paring  vector
$\vec{d}$
and the symmetry of the form factor. 
If $\mathrm{sign} (\Delta_k) = \mathrm{sign} (\Delta_{k+q}) $, the resonance appears in the plane perpendicular to the pairing direction. 
If $\mathrm{sign} (\Delta_k) = -\mathrm{sign} (\Delta_{k+q})$, the resonance appears in the susceptibility parallel to the pairing direction.

\section{Notation of Fourier transform} 
We define primitive vectors of the primitive unit cell as
\begin{eqnarray}
\bm{a}_1 &=& \frac{1}{2} (1,1,-1), \\
\bm{a}_2 &=& \frac{1}{2} (-1,1,1), \\
\bm{a}_3 &=& \frac{1}{2} (1,-1,1),
\end{eqnarray}
where there are two sublattices in the primitive unit cell. The corresponding reciprocal vectors are
\begin{eqnarray}
\bm{b}_1 &=& 2\pi (0,1,1) \\
\bm{b}_2 &=& 2\pi (1,0,1) \\
\bm{b}_3 &=& 2\pi (1,1,0).
\end{eqnarray}

We then discuss the two different definitions of Fourier transformation:
\ba 
&&c_{\bm{k},\alpha} = \sum_{\bm{R}} c_{i,\alpha}e^{-i\bm{k}\cdot \bm{R}_i  }, \nonumber\\ 
&&\tilde{c}_{\bm{k},\alpha} = \sum_{\bm{R}} c_{i,\alpha}e^{-i\bm{k}\cdot (\bm{R}_i+\bm{r}_\alpha)  }, \nonumber
\ea 
where $\alpha$ is the sublattice index, $\bm{R}_i$ is the position of $i$-th primitive unit cell and $\bm{r}_{\alpha}$ is the position of sublattice $\alpha$ in each unit cell [$\bm{r_A}=(0,0,\eta/2),\bm{r_B}=(0,0,-\eta/2)$]. $\tilde{c}_{\bm{k},\alpha}$ explicitly includes the additional phase factor caused by the sublattice displacement and two definitions are connected by the following relation
\ba 
{c}_{\bm{k},\alpha}  =\tilde{{c}}_{\bm{k},\alpha}e^{i\bm{k}\cdot \bm{r_\alpha} }. 
\ea 
Correspondingly, we can define the Fourier transformation of spin operators as 
\ba 
&&S^\mu_{\alpha}(\bm{q})= \sum_{\bm{R}} S^\mu_{i,\alpha}e^{-i\bm{q}\cdot \bm{R}_i  } = \sum_{\bm{k} }c_{\bm{k},\alpha}^\dag \sigma^\mu c_{\bm{k+q},\alpha}\nonumber,\\ 
&&\tilde{S}^\mu_{\alpha}(\bm{q}) = \sum_{\bm{R}} S^\mu_{i,\alpha}e^{-i\bm{q}\cdot (\bm{R}_i+\bm{r}_\alpha)  }= \sum_{\bm{k} }\tilde{c}_{\bm{k},\alpha}^\dag \sigma^\mu \tilde{c}_{\bm{k+q},\alpha}, \nonumber
\ea 
where $\mu=x,y,z$. And 
\ba 
\bm{S}_{\alpha}(\bm{q})  =\tilde{\bm{S}}_{\alpha}(\bm{q})e^{i\bm{q}\cdot \bm{r_\alpha} }. 
\ea
The corresponding spin susceptibilities are
\ba 
&&\chi^\mu_{\alpha\gamma}(\bm{q},i\omega) =\langle S^\mu_{\alpha}(\bm{q},i\omega)S^\mu_{\gamma}(-\bm{q},-i\omega) \rangle \nonumber \\
&&\tilde{\chi}^\mu_{\alpha\gamma}(\bm{q},i\omega) =\langle \tilde{S}^\mu_{\alpha}(\bm{q},i\omega)\tilde{S}^\mu_{\gamma}(-\bm{q},-i\omega) \rangle \nonumber,
\ea 
which are related to each other via 
\ba 
\chi^\mu_{\alpha\gamma}(\bm{q},i\omega) = \tilde{\chi}^\mu_{\alpha\gamma}(\bm{q},i\omega)e^{i\bm{q}\cdot(\bm{r}_\alpha-\bm{r}_\gamma)  }.
\label{eq:sus_transf}
\ea 

In the main text, we use the definition of $c_{\bm{k},\alpha}$ and $\bm{S}_{\alpha}(\bm{q})$ in order to simplify the notation. However, in the calculation of dynamical spin susceptibility, we transform back to the $\tilde{\chi}$ via Eq.~\ref{eq:sus_transf} and sum over all the scattering channels. The final spin susceptibilities
 are defined as
\ba 
\label{eq:fm_afm}
\chi^\mu(\bm{q},i\omega) &=& \sum_{\alpha\gamma}\tilde{\chi}^\mu_{\alpha\gamma}(\bm{q},i\omega) = \sum_{\alpha\gamma} \chi_{\alpha\gamma}^\mu(\bm{q},i\omega) e^{-i\bm{q}\cdot (\bm{r_\alpha}-\bm{r_\gamma}) }\nonumber \\
&=&\chi^\mu_{AA}(\bm{q},i\omega) + \chi^\mu_{BB}(\bm{q},i\omega) 
+\chi^{\mu}_{AB}(\bm{q},i\omega)e^{-i\bm{q}\cdot (\bm{r}_A - \bm{r}_B)} +\chi^{\mu}_{BA}(\bm{q},i\omega)e^{-i\bm{q}\cdot (-\bm{r}_A + \bm{r}_B)}. 
\ea
We note that for $q_z=0$, 
in Fig.~\ref{fig:res},
$\Tilde{\chi} = \chi$.

\section{Calculation of spin resonance}
We utilize the random phase approximation (RPA) to calculate the dynamical spin susceptibility 
of the spin-$x$ component
in the spin-triplet $\Delta_y$
pairing case. The corresponding equations in matrix form are
\begin{equation}
    \chi(\qbf,i\omega) =\chi^{0}(\qbf,i\omega) [\mathbb{I}+ J(\qbf)\chi^{0}(\qbf,i\omega)]^{-1} ,
\end{equation}
where $\chi^0$ is the bare susceptibility with the following matrix:
\begin{eqnarray}
    \chi^{0}(\qbf,i\omega) = \mx
    \chi^{0}_{AA}(\qbf,i\omega) & \chi^{0}_{AB}(\qbf,i\omega) \\
    \chi^{0}_{BA}(\qbf,i\omega) & \chi^{0}_{BB}(\qbf,i\omega)
    \ex.
\end{eqnarray} 

In practice, the dominant contributions to the low frequency spin susceptibility come from the intraband processes of the
 lowest band. Using Eq.~\ref{eq:project}, we approximately calculate bare spin susceptibility via
\begin{eqnarray}
    \chi^{0}(\qbf,i\omega) = \sum_{\kbf} \mx
    \chi^{0}_{aa}(\qbf,\kbf,i\omega) & u_{\kbf+\qbf}^*u_{\kbf} \chi^{0}_{aa}(\qbf,\kbf,i\omega)\\
   u_{\kbf+\qbf}u_{\kbf}^*\chi^{0}_{aa}(\qbf,\kbf,i\omega) & \chi^{0}_{aa}(\qbf,\kbf,i\omega)
    \ex,
\end{eqnarray}
where $\chi_{aa}^0$ denotes the contribution from the lowest bands and 
\begin{equation}
\chi_{aa}^{0}(\qbf,\kbf, i\omega) \sim \sum_k \frac{1}{2}\left(1 - \frac{\epsilon_-(\kbf)\epsilon_-(\kbf+\qbf)-d_1(\kbf)d_1(\kbf+\qbf)}{E_-(\kbf)E_-(\kbf+\qbf)} \right)\ \frac{1-f(E_-(\kbf))-f(E_-(\kbf+\qbf))}{-i\omega+E_-(\kbf)+E_-(\kbf+\qbf)}.
\end{equation} 
From the above equations,
we can calculate $\chi_{\alpha\gamma}(\bm{q},i\omega)$. In turn,
 the total spin susceptibility is given by $\chi(\bm{q},i\omega) = \sum_{\alpha\gamma} \chi_{\alpha\gamma}(\bm{q},i\omega) e^{-i\bm{q}\cdot (\bm{r_\alpha}-\bm{r_\gamma}) } $.

\begin{figure}[t!]
\centering
\includegraphics[width=0.8\columnwidth]{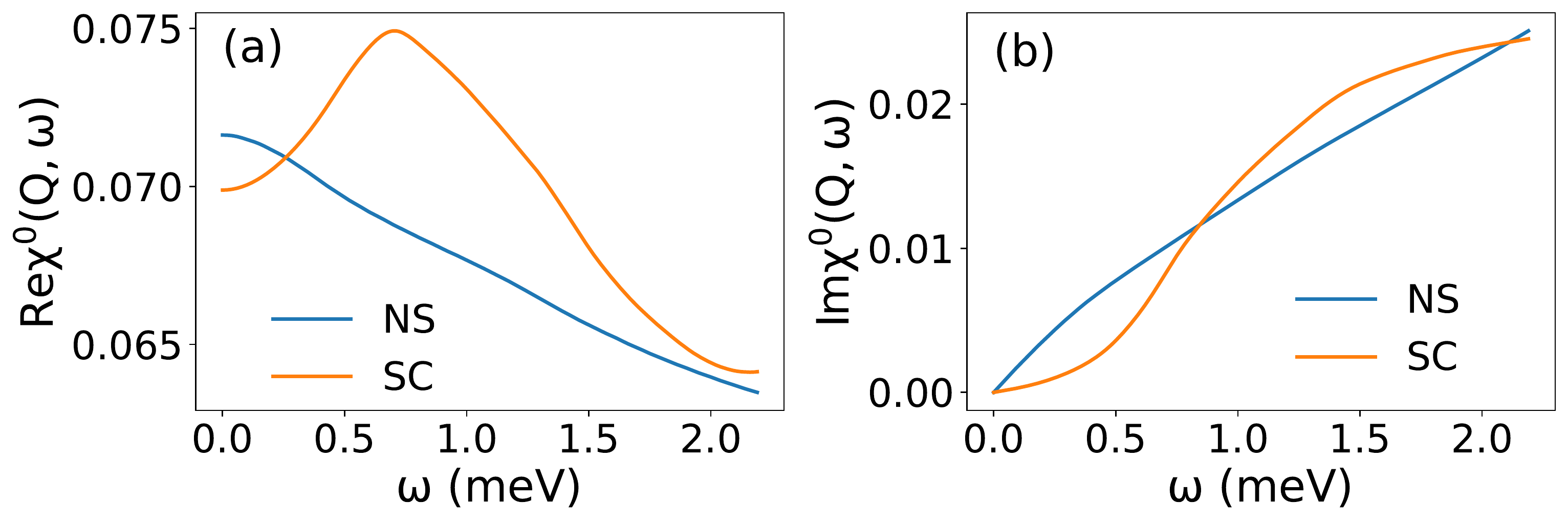}
\caption{ 
(a) The real part and (b) the imaginary part of the bare spin susceptibility in the normal state (NS) and superconducting state (SC). 
}
\label{fig:chi0}
\end{figure} 

\section{Bare susceptibilities}

Here, we present the bare susceptibility along $x$ direction

\begin{equation}
\chi^{0}(Q, i\omega) = \sum_{\alpha,\gamma} \chi^{0}_{\alpha\gamma}(Q,i\omega)e^{-iQ\cdot(\bm{r}_\alpha-\bm{r}_\gamma)}
\end{equation}

of both the normal state (NS) and the superconducting state (SC) in Fig.~\ref{fig:chi0}.

\section{magnetic exchange interactions inferred from
INS experiments}

As an application of
 Eq.~\ref{eq:fm_afm}, we consider the classical limit and suppose the magnetic form factor of individual $U$ atoms as $f_{\alpha}(\qbf)$ ($\alpha=A/B$) and suppose the coherent factor $u_{k}=1$ everywhere, which means $\chi_{AA}(\qbf) = \chi_{BB}(\qbf) = f^2(\qbf)$ and $\chi_{AB}(\qbf) = \chi_{BA}(\qbf) = f_A(\qbf)f_B(\qbf)$.
Eq.~\ref{eq:fm_afm} can be reorganized as 
\be
\chi^{\mu}(\qbf) = 2f^2(\qbf) + f_A(\qbf)f_B(\qbf)e^{i\qbf (\bm{r}_A - \bm{r}_B)} + f_A(\qbf)f_B(\qbf)e^{-i\qbf (\bm{r}_A - \bm{r}_B)}.
\ee
If the magnetic correlation between $A$ and $B$ sublattices within the same unit cell is ferromagnetic, i.e. $ f_A(\qbf)=f_B(\qbf) = f(\qbf)$, then $\chi^{\mu}_{AB} = \chi^{\mu}_{AB}>0$ and we have
\be
\begin{aligned}
\chi^{\mu}(\qbf) &= f^2(\qbf) \left( 2+2\cos q_z \eta \right) \\
& = 4 f^2(\qbf) \cos^2 \frac{q_z \eta}{2}.
\label{eq:fm}
\end{aligned}
\ee
For the antiferromagnetic correlation($ f_A(\qbf)=-f_B(\qbf) = f(\qbf)$), $\chi^{\mu}_{AB} = \chi^{\mu}_{AB} < 0$, and we have 
\be
\begin{aligned}
\chi^{\mu}(\qbf) &= f^2(\qbf) \left( 2-2\cos  q_z \eta \right) \\
& = 4 f^2(\qbf) \sin^2 \frac{\ q_z \eta}{2}.
\label{eq:afm}
\end{aligned}
\ee
Following Eqs.~\ref{eq:fm},~\ref{eq:afm}, the inelastic neutron-scattering results imply that 
the correlation between $A$ and $B$ sublattices are FM,
thereby suggesting a ferromagnetic $I$ \cite{Knafo2021}.

Further experimental constraints, on $J_{r \ge 2}$, are less direct, and are primarily based on the wavevector for the observed
AF correlations.
The observed $Q_y \gtrsim \pi$ suggests that both $J_4$ and $J_2$ are AF. Likewhise, the observed $Q_x=0$ suggests that $J_3$ is FM.
\end{document}